\documentclass[sigconf,letterpaper]{acmart}

\AtBeginDocument{%
  \providecommand\BibTeX{{%
    \normalfont B\kern-0.5em{\scshape i\kern-0.25em b}\kern-0.8em\TeX}}}

\setcopyright{acmcopyright}
\copyrightyear{2022}
\acmYear{2022}
\acmDOI{}

\acmConference[ICSE 2022]{The 44th International Conference on Software Engineering}{May 21–29, 2022}{Pittsburgh, PA, USA}

\usepackage{algorithmic}
\usepackage{graphicx}
\usepackage{textcomp}
\usepackage{xcolor}
\usepackage{url}
\usepackage{hyperref}
\usepackage{multirow}
\usepackage{mathpazo}
\usepackage{setspace}
\usepackage{hyperref}
\usepackage{caption}
\usepackage{subcaption}
\usepackage{wrapfig}
\usepackage{enumitem}
\usepackage{booktabs}
\usepackage{flafter}
\usepackage{xspace}
\usepackage{lipsum}

\newcommand{\ie}{\emph{i.e.,}\xspace}
\newcommand{\eg}{\emph{e.g.,}\xspace}

\newcommand{\etal}{\emph{et~al.}\xspace}
\newcommand{\secref}[1]{Section~\ref{#1}\xspace}

\newcommand{\figref}[1]{Figure~\ref{#1}\xspace}

\newcommand{\tabref}[1]{Table~\ref{#1}\xspace}
\newcommand{\tool}[1]{{\sc #1}\xspace}
\newcommand{\metricity}{\tool{M3triCity}}
\newcommand{\codecity}{\tool{CodeCity}}

\newcommand{\urltt}[1]{\textbf{\texttt{\url{#1}}}}
\newcommand{\seeurl}[1]{\footnote{See~\urltt{#1}}}

\usepackage{tikz}
\newcommand*\circled[1]{\tikz[baseline=(char.base)]{
            \node[shape=circle,draw,inner sep=0pt] (char) {#1};}}
            
\copyrightyear{2022} 
\acmYear{2022} 
\setcopyright{acmcopyright}\acmConference[ICSE '22 Companion]{44th International Conference on Software Engineering Companion}{May 21--29, 2022}{Pittsburgh, PA, USA}
\acmBooktitle{44th International Conference on Software Engineering Companion (ICSE '22 Companion), May 21--29, 2022, Pittsburgh, PA, USA}
\acmPrice{15.00}
\acmDOI{10.1145/3510454.3516831}
\acmISBN{978-1-4503-9223-5/22/05}

\begin{document}

\title{M3triCity: Visualizing Evolving Software \& Data Cities}

\author{Susanna Ardig\`o, Csaba Nagy, Roberto Minelli, Michele Lanza}
\affiliation{%
  \institution{REVEAL @ Software Institute -- USI, Lugano}
  \city{}
  \country{}
}
\renewcommand{\shortauthors}{Ardigò \etal}


\begin{abstract}

The city metaphor for visualizing software systems in 3D has been widely explored and has led to many diverse implementations and approaches. Common among all approaches is a focus on the software artifacts, while the aspects pertaining to the data and information (stored both in databases and files) used by a system are seldom taken into account.

We present \metricity, an interactive web application whose goal is to visualize object-oriented software systems, their evolution, and the way they access data and information. We illustrate how it can be used for program comprehension and evolution analysis of data-intensive software systems.

\textbf{Demo video URL:} \urltt{https://youtu.be/uBMvZFIlWtk}

\end{abstract}




\maketitle


\section{Introduction}

Program comprehension is a fundamental activity for software maintenance and evolution. Developers spend considerably more time reading and understanding existing code rather than writing new code \cite{Minelli2015know}. Software visualization is a popular technique to perform program comprehension~\cite{Stas1998SoftViz}. Many techniques have been proposed, ranging from simple 2D displays, such as polymetric views \cite{Lanza2003a} and UML diagrams to more complex 3D techniques, even extending into the realm of virtual (VR) and augmented reality (AR)~\cite{VRcities, morenocodecity,Merino18AR}.

We present \metricity a web application that visualizes software systems in 3D, focusing on the evolution of systems and how they use and access data~\cite{Pfah2020a, Ardi2021a}. \metricity leverages the city metaphor~\cite{Knight2000, Panas2003, Panas2007, Steinbrueckner2010, ExplorViz, ViDI} in the vein of \codecity \cite{Wettel2008} and runs on any modern web browser.

\section{M3triCity in a Nutshell}

\begin{figure}[ht]
\centering
\includegraphics[width=0.7\columnwidth]{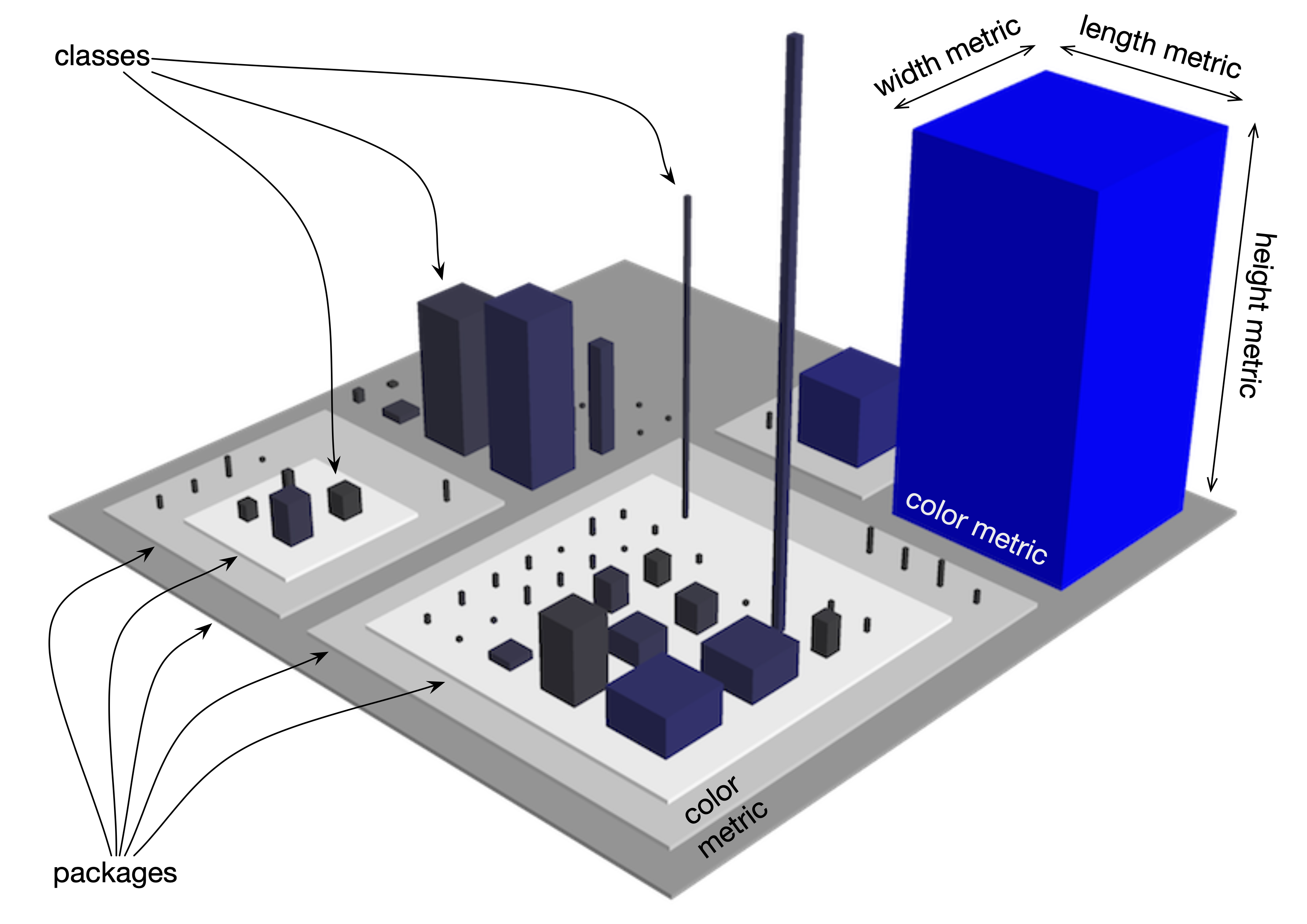}
\caption[\codecity]{CodeCity and the City Metaphor}\label{fig:CodeCity}
\end{figure}

In a nutshell, \metricity is a web-based evolution of \codecity, which popularized the city-based visualization of software systems through the city metaphor~\cite{Wettel2007} (See \figref{fig:CodeCity}).

\begin{figure}[ht]
    \centering
    \includegraphics[width=0.75\columnwidth]{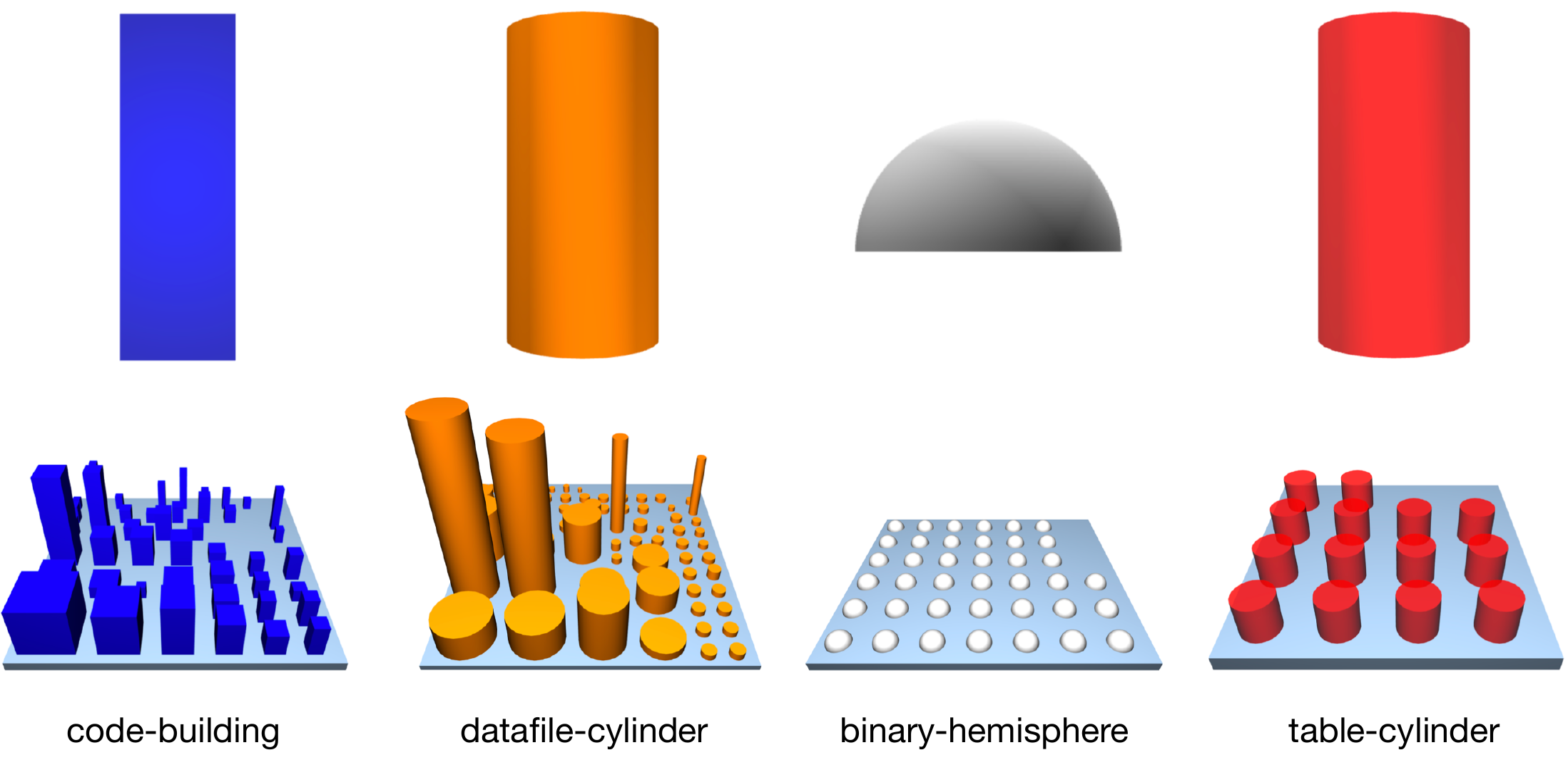}
    \caption{The Glyphs Used by M3triCity}\label{fig:meshes}
\end{figure}

In \codecity every class is visualized as a building with metrics mapped onto the base, height, and color, while packages were visualized as nested districts. \metricity expands on it through a number of features and concepts, namely: (i) it distinguishes between the file types and uses different glyphs (\ie depictions) for them, as we see in \figref{fig:meshes}.

Furthermore, (ii) it takes the evolution of a system into account for the layout of the city structure, as we detail in \secref{usage}; (iii) it infers and visualizes the databases used by a system, as we detail in \secref{modeling}, and (iv) it provides higher accessibility by being publicly available as a web application available at \urltt{https://metricity.si.usi.ch/v2}.

\begin{figure*}[ht]
    \centering
    \includegraphics[width=0.88\textwidth]{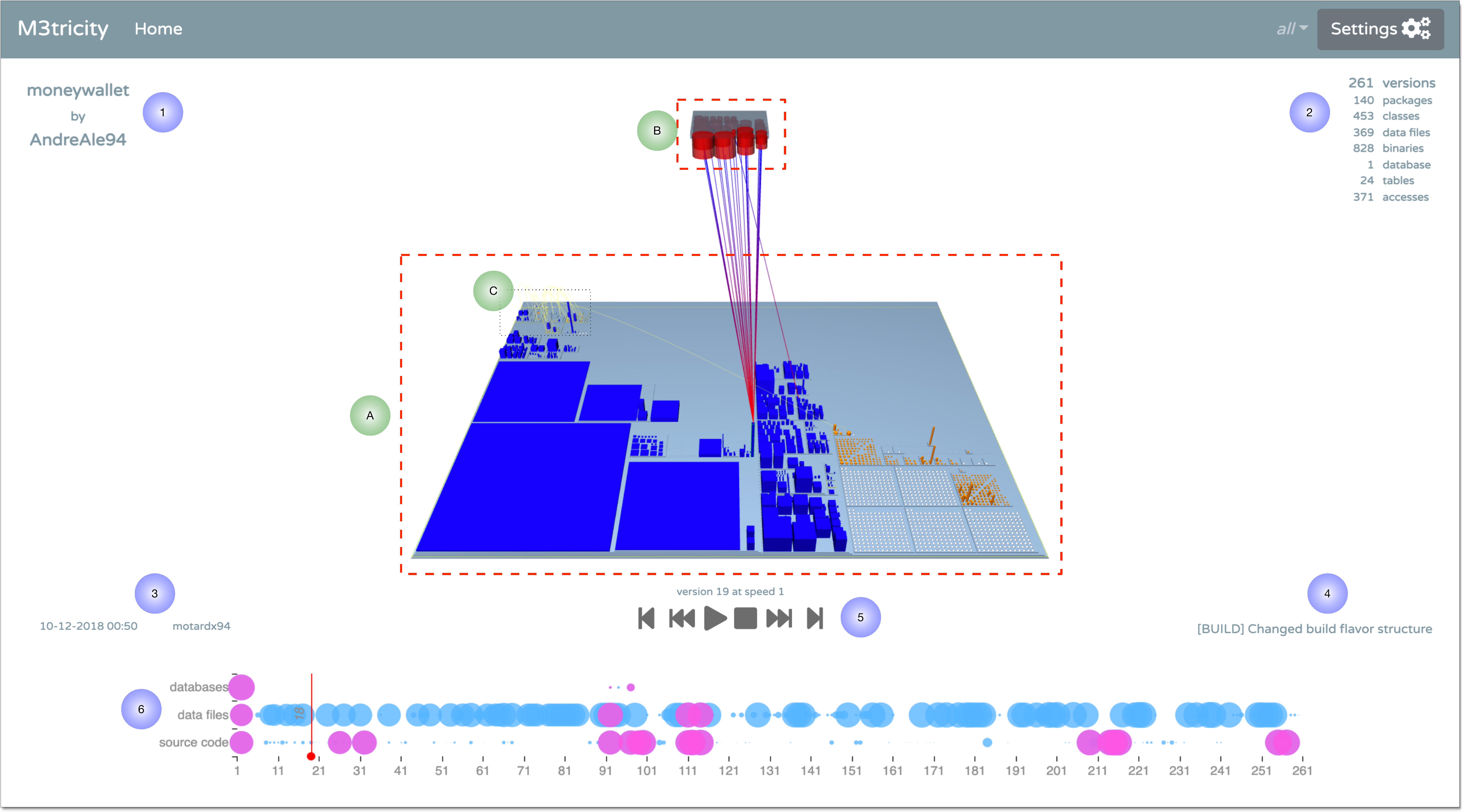}
    \caption{The Main Page of M3triCity}
	\label{fig:metricity2-example}
\end{figure*}

\begin{figure*}[ht]
    \centering
    \includegraphics[width=0.75\textwidth]{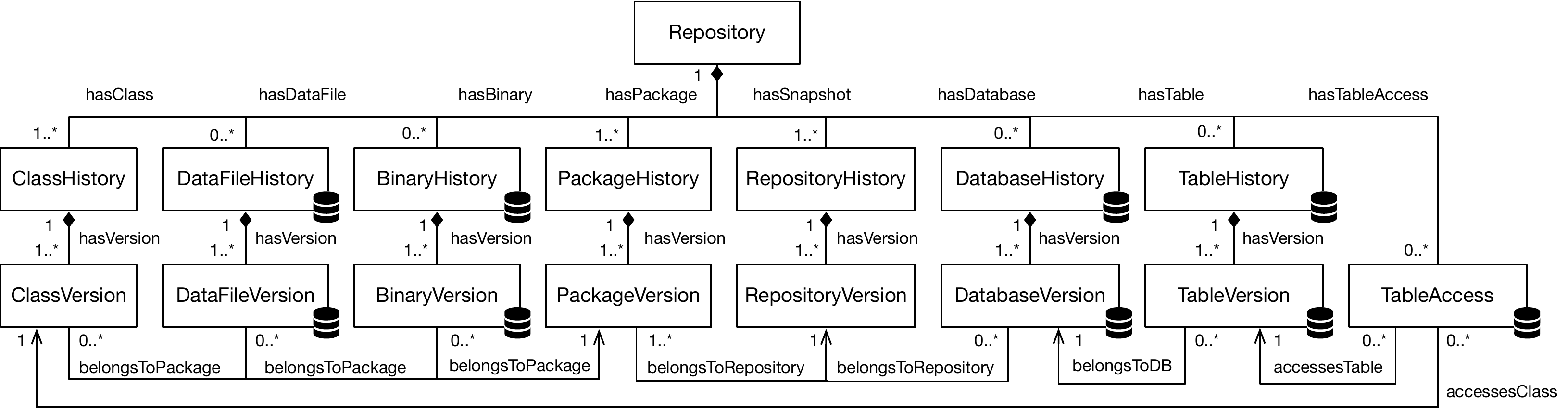}
    \caption{The Evolution Model of M3triCity}\label{fig:model}
    \vspace{-2mm}
\end{figure*}

\subsection{The User Interface of M3triCity}

\figref{fig:metricity2-example} shows the main user interface of \metricity. At the center we see the software city \circled{A}, with folders and files represented as buildings and nested districts. Above the city the sky is used to visualize the (inferred) database(s) and their tables \circled{B}, where connecting lines represent the accesses performed from the source code.

Various information about the system being visualized \circled{1} is also displayed \circled{2}. As \metricity is geared towards evolution comprehension, additional panels provide information about the currently visualized commit \circled{3}\circled{4}. To facilitate moving through time \metricity provides a control panel \circled{5} to easily moving forward/backward between different commits as well as fast back/forwarding and pausing. A timeline at the bottom \circled{6} provides a global overview of the system evolution with additional details pertaining to the commits (\ie along the timeline), as well as instantaneous access to a specific place in the commit history. The whole city can be rotated, the user can also change its point of view and zoom in and out. Structural changes (\ie moving of entities) are depicted using yellow curved arcs (see top left annotation with black dots \circled{C}). When the user clicks on an artifact, it is highlighted both in the main visualization as well as along the timeline, denoting all commits in which the entity was involved. More customizations are also available in the settings panel.

\subsection{Modeling Evolving Data-Intensive Systems} \label{modeling}

\figref{fig:model} depicts the meta-model of \metricity. Evolving software artifacts are modeled using ``histories'' in the vein of Girba's evolution meta-model \cite{GirbaThesis}. For each artifact history, we model each version including binary files and data files (\eg JSON, XML). Databases are inferred through \tool{SQLInspect}~\cite{SQLInspectICSEDemo}. For each entity, \metricity computes various metrics, summarized in \tabref{tab:supportedMetrics}.

\begin{table}[!ht]
\vspace{-1mm}
\centering
\footnotesize
\begin{tabular}{ll|ll}
\toprule
\textbf{Entity}		& \textbf{Metric Name}			& \textbf{Entity}		& \textbf{Metric Name}	\\ \midrule
\multirow{4}{*}{Class}		& \# Instance Variables	& \multirow{4}{*}{Data File}	& \# Entities			\\
							& \# For Loops			& 								& \# Entity Types				\\
							& \# Methods				&								& Max \# Properties per Entity	\\
							& \# Lines of Code		&								& Max Nesting Level			\\ \cmidrule{1-4}
\multirow{2}{*}{Table}		& \# Columns				& 	Binary						& Size 				\\
							& \# Table Accesses	&&\\		
\bottomrule

\end{tabular}	
\captionof{table}{The Metrics Supported by M3triCity}\label{tab:supportedMetrics}
\end{table}

\subsection{Architecture}

\begin{figure}[ht]
    \centering
    \includegraphics[width=0.75\columnwidth]{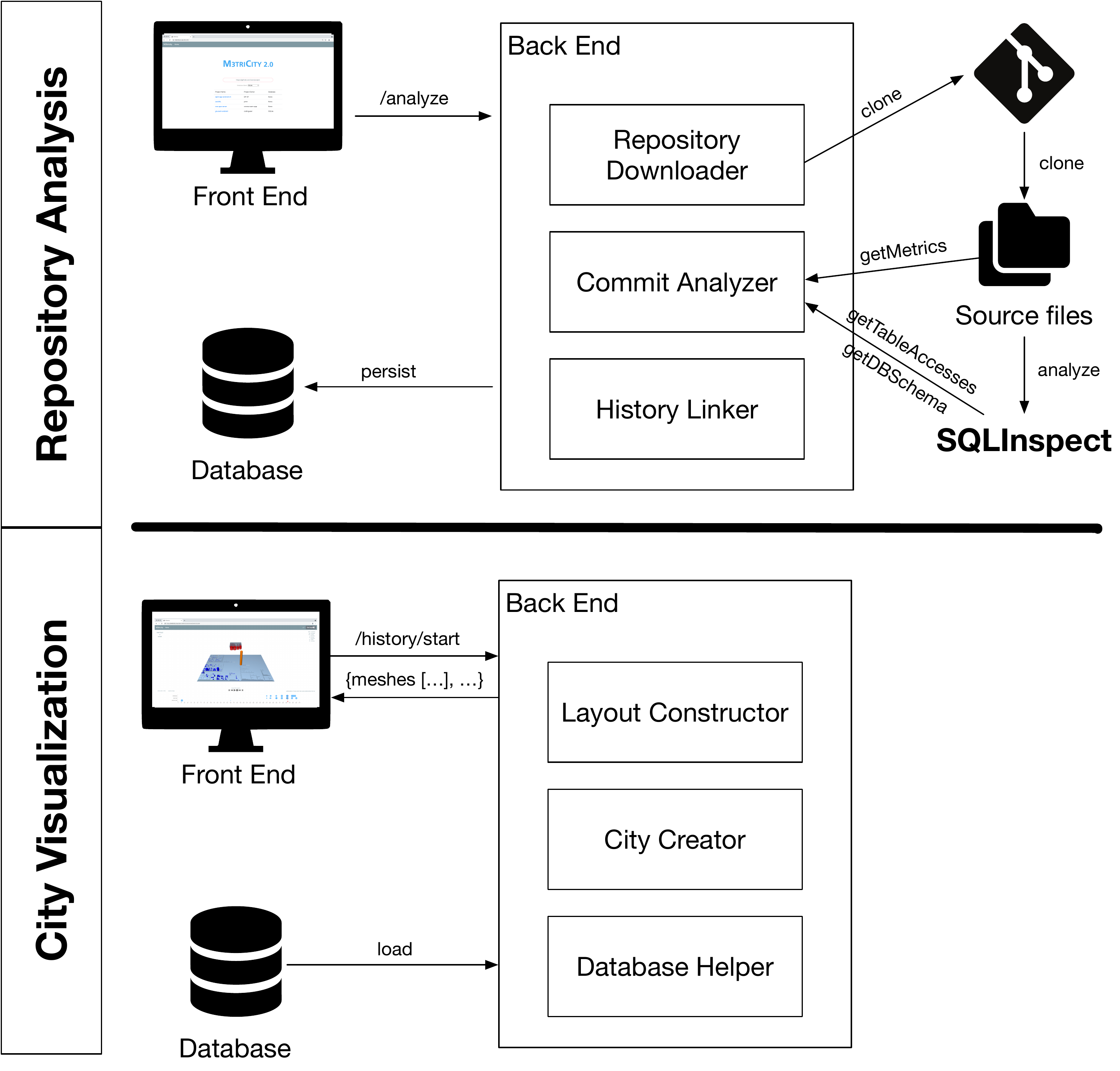}
    \caption{The Architecture of M3triCity}\label{fig:repository_analysis}
\end{figure}

\figref{fig:repository_analysis} shows the architecture of \metricity.

The frontend of \metricity is implemented in TypeScript.\seeurl{https://www.typescriptlang.org} It uses Vue.js\seeurl{https://vuejs.org} for the user interface. The 3D visualization is created using Babylon.js.\seeurl{https://www.babylonjs.com} The backend is a Spring Boot \seeurl{https://spring.io/projects/spring-boot} application implemented using Java 11\seeurl{https://docs.oracle.com/en/java/javase/11/docs/api/} and Gradle 6.\seeurl{https://docs.gradle.org/6.3/release-notes.html}

To start the analysis, the user indicates the URL of a repository and, as an optional parameter, the database type. The frontend contacts the backend through the public REST API endpoint \textit{analyze}. The execution starts in the module \textit{Repository Downloader} which contacts git to clone the repository. The \textit{Commit Analyzer} module iterates through all files of each snapshot of the given repository in chronological order, classifies them, analyzes them and then extracts the metrics. The project uses \tool{SQLInspect}~\cite{SQLInspectICSEDemo} to reverse engineer the schema of the database and the interactions with the source code. The histories of the entities are created by linking the versions. At last, all information is persisted in a MongoDB database.

\subsection{Usage} \label{usage}

The user starts the visualization of a city by selecting a processed repository. The repository-related information is loaded from the database with the \textit{Database Helper} component. The computation of the city layout is handled by the \textit{Layout Constructor} component which iterates through all the entities of the city. \metricity considers evolution as a first-class citizen, implementing a history-resistant layout, which allocates a dedicated position to each artifact throughout the lifetime~\cite{Pfah2020a}. 
The \textit{City Creator} module creates the meshes information that are then sent to the frontend which renders them as 3D meshes.

\section{Software City Tales}

\begin{figure}[ht]
\centering
\begin{subfigure}[t]{0.45\columnwidth}
\centering
\includegraphics[width=\textwidth]{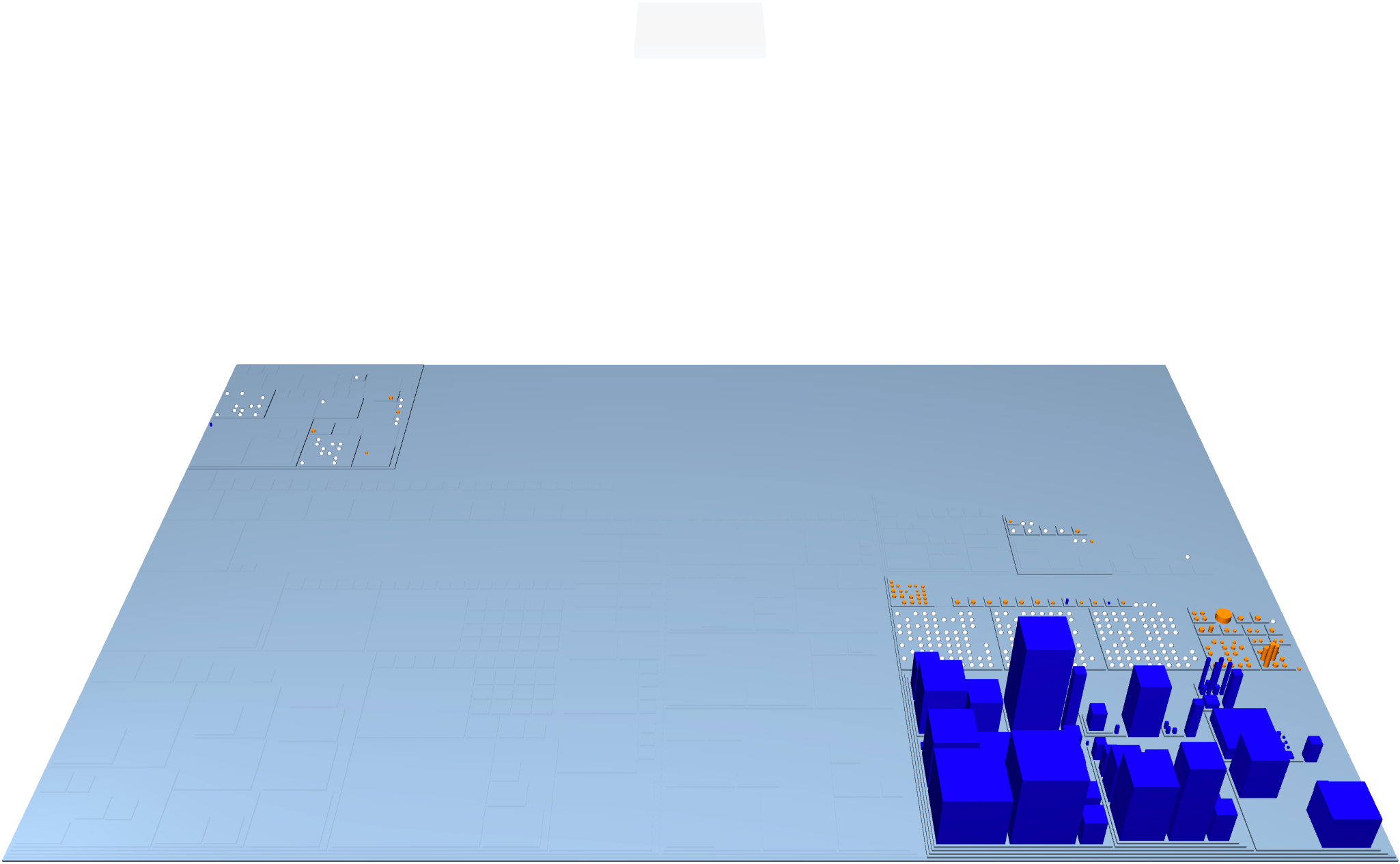}
\caption{13 May 2012 at 19:27} \label{fig:gnucash-0001}
\end{subfigure}
\hfil
\begin{subfigure}[t]{0.45\columnwidth}
\centering
\includegraphics[width=\textwidth]{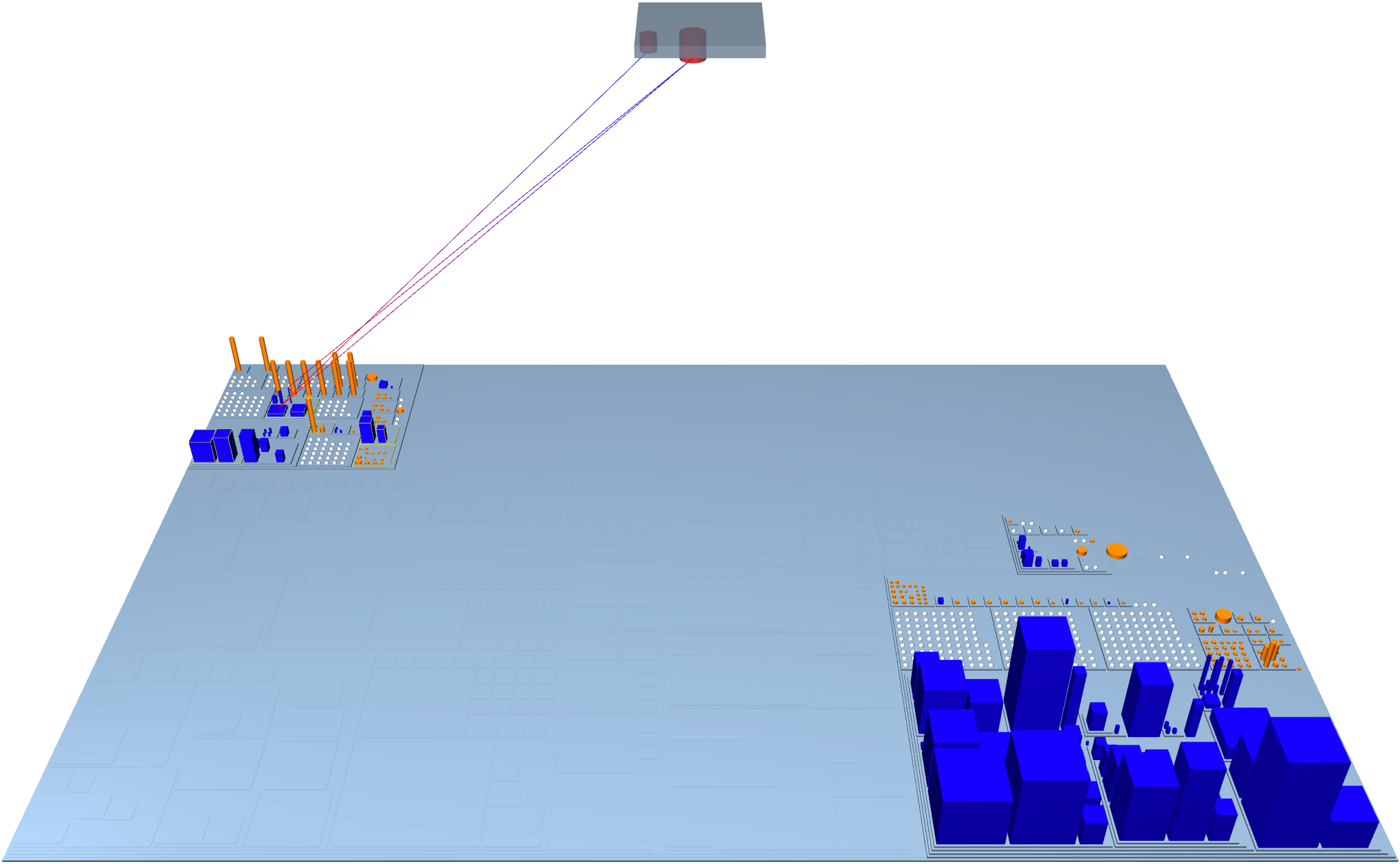}
\caption{4 November 2012 at 17:20} \label{fig:gnucash-108}
\end{subfigure}
\hfil
\begin{subfigure}[t]{0.45\columnwidth}
\centering
\includegraphics[width=\textwidth]{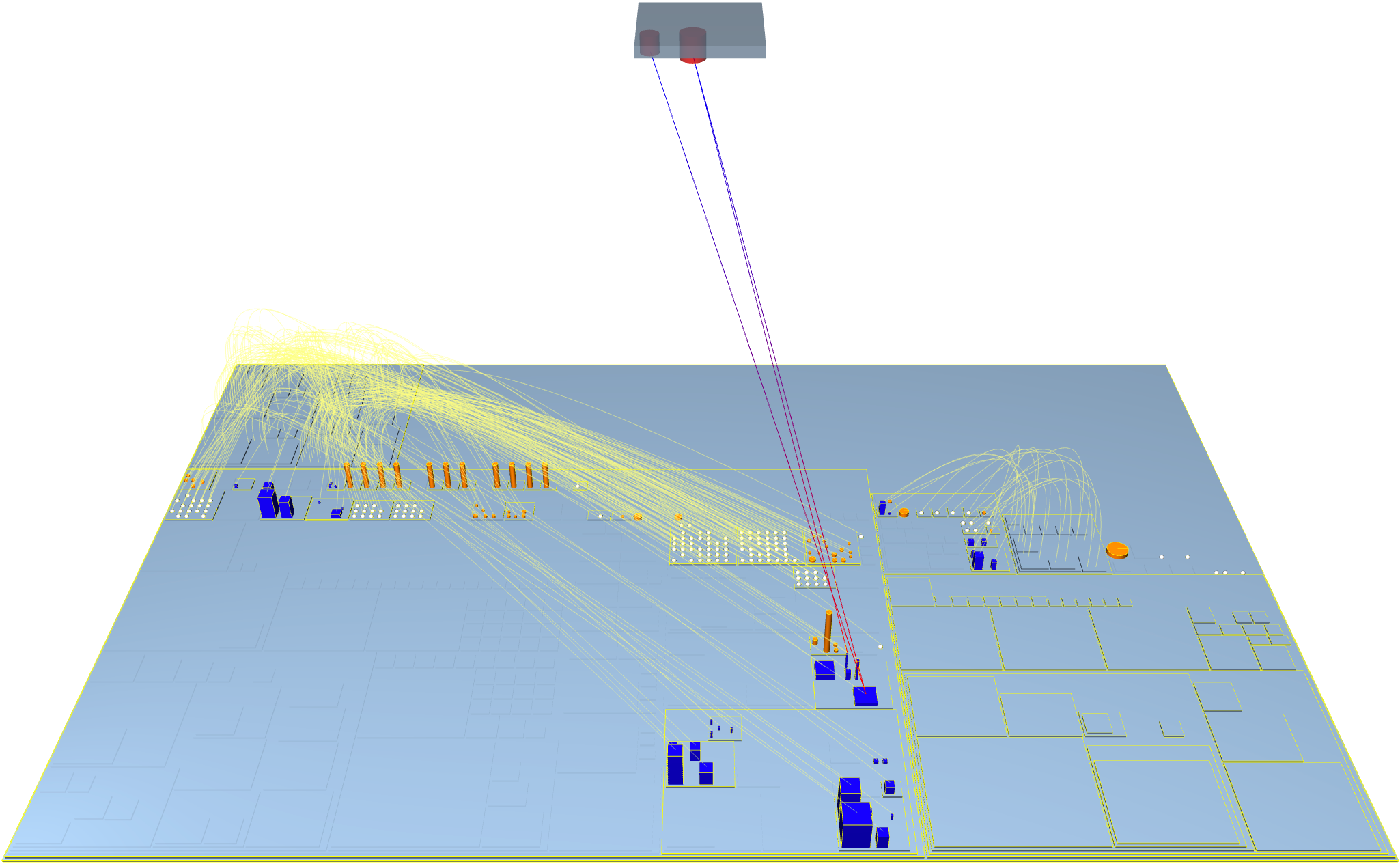}
\caption{31 January 2013 at 00:29} \label{fig:gnucash-181}
\end{subfigure}
\hfil
\begin{subfigure}[t]{0.45\columnwidth}
\centering
\includegraphics[width=\textwidth]{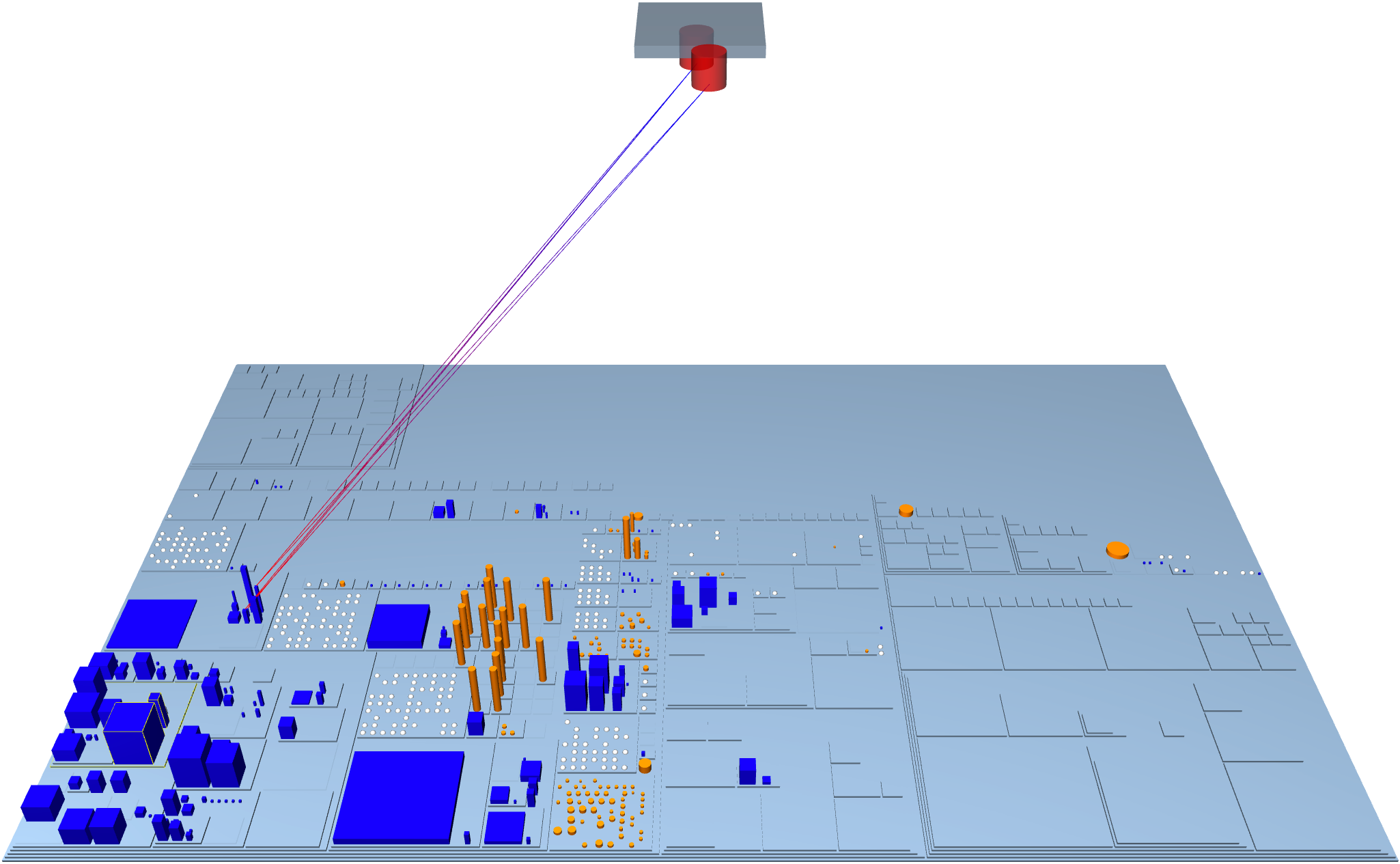}
\caption{18 September 2015 at 19:06} \label{fig:gnucash-1050}
\end{subfigure}
\hfil
\begin{subfigure}[t]{0.45\columnwidth}
\centering
\includegraphics[width=\textwidth]{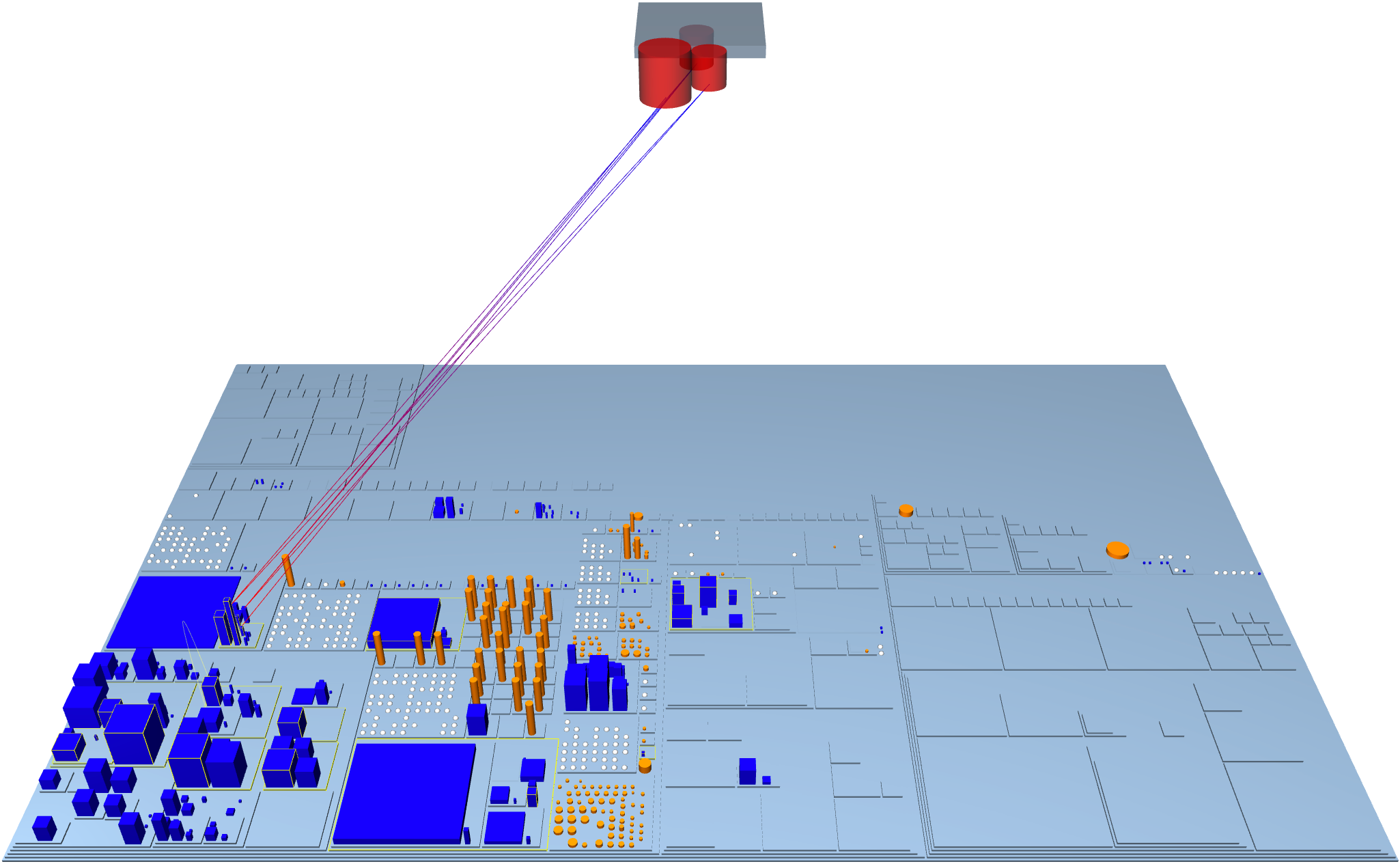}
\caption{28 December 2015 at 10:06} \label{fig:gnucash-1251}
\end{subfigure}
\hfil
\begin{subfigure}[t]{0.45\columnwidth}
\centering
\includegraphics[width=\textwidth]{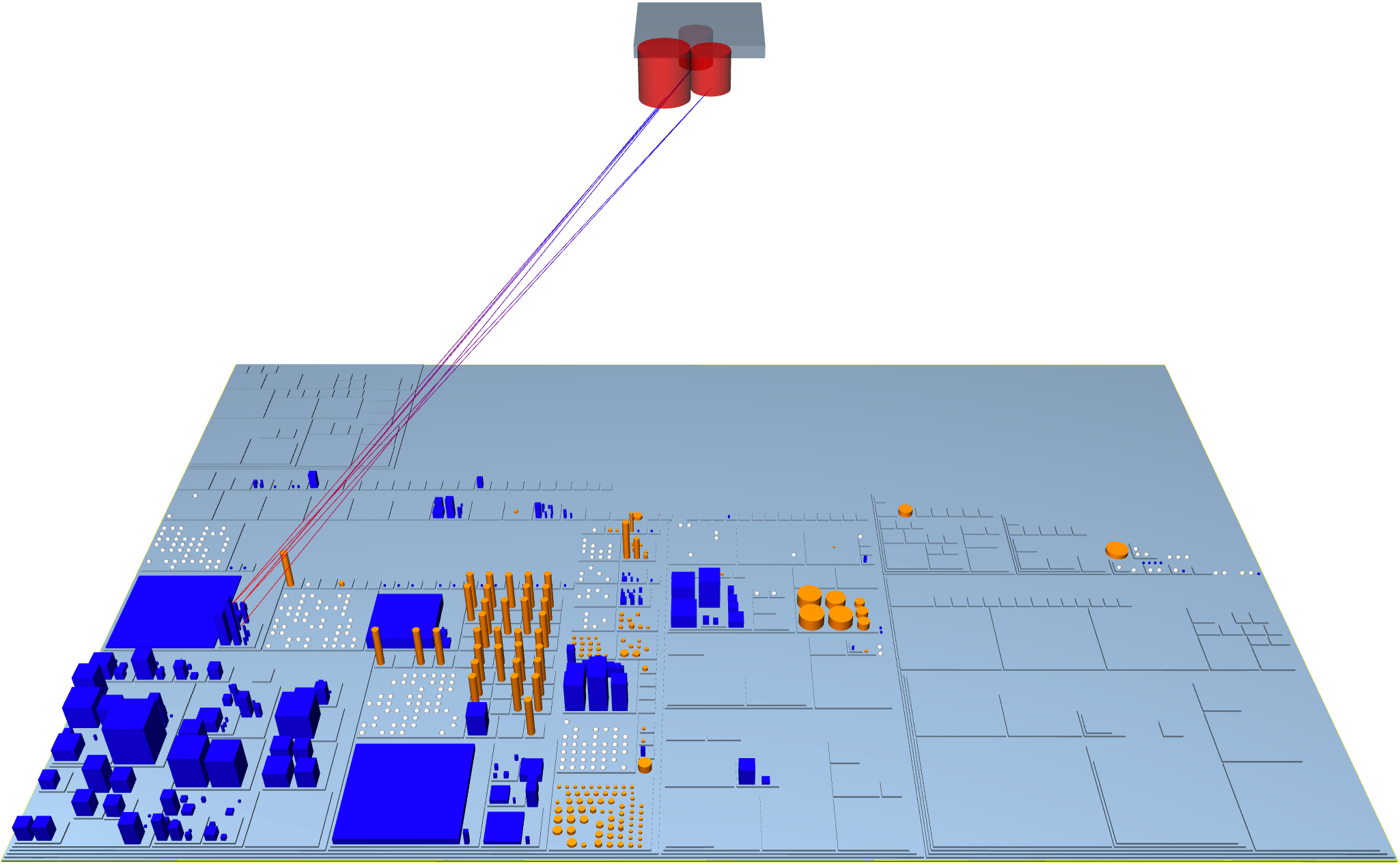}
\caption{2 December 2020 at 08:13} \label{fig:gnucash-1727}
\end{subfigure}
\caption{The Evolution of GnuCash-Android}
\label{fig:gnucash}
\end{figure}

We illustrate how \metricity can be used to comprehend the evolution of a system, by using the \tool{GnuCash-Android}\seeurl{https://github.com/codinguser/gnucash-android} Android companion app of the \tool{GnuCash} accounting program as an example. \tool{GnuCash-Android} allows recording transactions on-the-go to import the data into \tool{GnuCash} later. The main branch of the project is composed of 1,730 commits by 46 contributors. 
\figref{fig:gnucash} shows six \metricity snapshots in the overall evolution.

{\bf May 13 2012: GnuCash-Android is born (\figref{fig:gnucash-0001}).} On May 13, 2012, Ngewi Fet creates the project repository with 83 Java classes, 85 data files, and 243 binaries. The source code is mainly located in the \texttt{com\_actionbarsherlock} package, nicely divided into sub-packages. The \texttt{res} folder contains three districts of images and several districts of data files. Separately, the module \textbf{GnucashMobile} has one Java class, four data files, and some binaries.

{\bf Nov 4 2012: The database is created (\figref{fig:gnucash-108}).} The developers added a database which is being used by a part of the system that has been added in the meantime. Test classes are growing as well. Data files have been added, mostly related to text constants that need to be displayed. 

{\bf Jan 31 2013: GnuCash-Android undergoes a major restructuring (\figref{fig:gnucash-181}).} Almost 450 files are deleted, and 220 are moved with the renaming of the folder \texttt{GnucashMobile} to \texttt{app}. The database-related classes are still present but the accesses to the tables are removed.

{\bf Sep 18 2015: Controlled evolution (\figref{fig:gnucash-1050}).} Fast forwarding 2 years, the system keeps evolving: the test suite is expanded, the developers started working on a user interface module. The database co-evolves with the system, with the addition and quick deletion of tables.

{\bf Dec 28 2015: Extending the tests (\figref{fig:gnucash-1251}).} The system evolves mostly with new tests. 

{\bf Dec 2 2020: Fast forward (\figref{fig:gnucash-1727}).} Five years later the system has grown considerably, with some classes reaching considerable size in terms of variables and methods. Data file districts have been added, complementing systematic database accesses on a well-organized DB featuring the three major tables \texttt{transactions}, \texttt{splits}, and \texttt{scheduled\_actions}.


\section{Related Work}

Since the seminal works of Reiss \cite{Reiss1995} and Young \& Munro \cite{Young1998}, many approaches to visualize software systems in 3D have been explored. The cities metaphor has been widely used and led to diverse implementations, such as the \tool{Software World} by Knight \etal \cite{Knight2000}, the visualization of communicating architectures by Panas \etal \cite{Panas2003}, \tool{Verso} by Langelier \etal \cite{Langelier2005}, \tool{CodeCity} by Wettel \etal \cite{Wettel2007, Wettel2008}, \tool{EVO-STREETS} by Steinbr\"{u}ckner \& Lewerentz \cite{Steinbrueckner2010}, \tool{CodeMetropolis} by Balogh \& Beszedes \cite{Balogh2013}, and \tool{VR City} by Vincur \etal \cite{Vincur2017}.

Only a few approaches considered presenting data(bases) together with the source code, mostly using the city metaphor. Meurice and Cleve presented \tool{DAHLIA} to visualize database schema evolution \cite{Meurice2016}, which uses the city metaphor where buildings in the city represent database tables. Zirkelbach and Hasselbring presented \tool{RACCOON} \cite{Zirkelbach2019}, which uses the 3D city metaphor to show the structure of a database based on entity-relationship diagrams. Marinescu presented a meta-model containing object-oriented entities, relational entities and object-relational interactions \cite{Marinescu2019}. \metricity does not separate source code and data(bases), like existing approaches, but it shows them with their interactions in the same city.


\section{Conclusion}

\metricity extends the original city metaphor by considering an important aspect that has been ignored up to now: the data.

Our tool visualizes object-oriented software systems, their evolution, and the way they access data and information. \metricity expands the original city metaphor by adding a number of features and concepts: using different glyphs to distinguish between the various file types, taking software evolution into account to layout the city, inferring and visualizing the databases used by a system, and providing higher accessibility by being publicly available as a web application.

To demonstrate the usefulness of our approach, we illustrate how \metricity can be used to comprehend the evolution of a data-intensive system: \tool{GnuCash-Android}.

\begin{acks}
We acknowledge the financial support of the Swiss National Science Foundation and the Fonds de la Recherche Scientifique for the project ``INSTINCT'' (SNF Project No. 190113).
\end{acks}

\bibliographystyle{ACM-Reference-Format}
\bibliography{biblio}

\end{document}